\documentclass[preprint,12pt]{elsarticle}
\usepackage{graphicx}
\usepackage{amssymb}
\usepackage{amsmath}
\usepackage{tabularx}
\usepackage{caption}
\usepackage{epstopdf}
\usepackage{amsfonts}
\usepackage{amsmath}

\journal{Nuclear Physics A}

\begin{document}

\begin{frontmatter}

\title{Approaching Pomeranchuk Instabilities from Ordered Phase: A Crossing-symmetric Equation Method} 

\author[a]{Kelly Reidy}
\author[a]{Khandker Quader\corref{cor1}}
\author[b]{Kevin Bedell}
\cortext[cor1]{Corresponding author: Khandker Quader, quader@kent.edu}
\address[a]{Department of Physics, Kent State University, Kent, OH 44242, USA}
\address[b]{Department of Physics, Boston College, Cestnut Hill, MA, USA}

\begin{abstract}
We explore features of a 3D Fermi liquid near generalized Pomeranchuk instabilities using a tractable crossing symmetric equation method. We approach the instabilities from the ordered ferromagnetic phase. We find ``quantum multi-criticality" 
as approach to the ferromagnetic instability drives instability in other channel(s). It is found that a charge nematic instability precedes and is driven by Pomeranchuk instabilities in both the $\ell=0$ spin and density channels. 
\end{abstract}


\end{frontmatter}

\section{Dedication} 
\label{sec:ded} 
\vspace{-0.5em}
(\small{by {Khandker Quader)}\\

\noindent
``{\it We are like dwarfs on the shoulders of giants. \\  
 So that we can see more than they, \\
 And things at a greater distance,\\
 Not by virtue of any sharpness of sight on our part,\\
 Or any physical distinction,\\
 But because we are carried high, \\
 And raised up by their giant size}" \\
  \noindent
  \small{(Bernard of Chartres/John Salisbury (12th century); Isaac Newton (17th century))}\\[-1em]

Much of the many-fermion physics that treats short-range underlying interaction and longer-range
quantum fluctuations on the same footing are rooted in the bold, seminal ideas of Gerry that some of
us, as his students, had the great fortune of learning first-hand from him. Over the years, Gerry's ``induced interaction" edifice 
gave us the confidence and guidance to build this into a tractable crossing-symmetric theory, and apply 
to interacting fermion problems in condensed matter and nuclear physics. This contribution is dedicated to the memory of Gerry!

\section{Introduction}
\label{sec:intro}

This work aims to present a general study of the physics near Pomeranchuk instabilities (PI)~\cite{Pomeran58} in 3D isotropic Fermi liquids~\cite{Landau56, Landau57}, when approached from the ordered side. Pomeranchuk instabilities (PIs) are instabilities of the Fermi liquid (FL) that occur when $F_\ell ^{s,a}\leq-(2\ell+1)$, where  $F_\ell ^{s,a}$  are the Landau Fermi liquid (FL) interaction functions. These are driven by forward scattering interactions and result in symmetry-breaking deformations of the 
Fermi surface (FS)~\cite{Pomeran58}. As a PI is approached, the susceptibility in the relevant channel (henceforth referred to as the ``critical channel")  will diverge, indicating a ``softness" of the Fermi surface with respect to its deformation. A familiar example of a Pomeranchuk instability is the ferromagnetic (Stoner) instability $F_0^a\to-1$; here, the Fermi surface splits into a spin-up surface and a spin-down surface, magnetic susceptibility diverges, and time-reversal symmetry is broken. Likewise,  $F_0^s \rightarrow -1$ marks an approach to a charge or density instability. In parameter space, PIs can be considered quantum critical points~\cite{SachdevBook}. 

A key issue is whether susceptibilities in ``non-critical" channels are affected when a PI is approached in some critical channel. One method for examining the behavior of systems around QCPs is using Hertz-Millis-type effective theories \cite{Hertz, Millis}; applying these theories, one finds that when a PI is approached in one channel, the effective mass, and hence $F_{1}^{s}$, also diverge. 
 A recent analysis~\cite{ChubukovCharge} of the properties of a 2D Fermi system in the paramagnetic state 
 near a charge nematic ($\ell = 2$) PI was done in terms of Landau FL theory. It was found that near the transition, the system enters into a new critical FL regime, in which all spin components of the FL interaction functions ($F_{\ell}^{a}$) and all charge components ($F_{\ell}^{s}$) with $\ell\neq2$ diverge at the critical point, while the $\ell=2$ charge FL component, $F_{2}^{s}\rightarrow-5$, the PI for this channel. However, owing to cancellation between divergent effective mass and divergent effective Landau component, non-critical channels susceptibilities were found not to be affected. Another work by the same authors \cite{ChubukovSpin} finds that a 2D FL crosses the $\ell=1$ instability in the spin channel before getting to the FM instability and all other possible instabilities near the FM QCP. 

In this work, we employ the tractable crossing symmetric equation (TCSE) method \cite{JacksonLandeSmith, BabuBrown, BickersScalapino, QuaderLittlePaper, AB87, ABBQ} to study approach to PIs from the ordered side, such as ferromagnetic phase.
 The TCSE method is a diagrammatic many-particle method that is used to calculate the Fermi liquid interaction functions through considering the s (particle-particle), t (particle-hole), and u (exchange particle-hole) channels in a conserving, self-consistent fashion. After partial resummation of diagrams in these channels, quasiparticle renormalization, and enforcement of crossing symmetry, one arrives at a set of coupled non-linear integral equations from which the FL interaction functions can be calculated. A unique aspect of this method is its ability to simultaneously consider underlying interactions of arbitrary strength and range,  and competing quantum fluctuations (density, spin, current, spin current, etc). Another aspect is that density and spin fluctuations, as well as higher-order fluctuations (such as current or spin current fluctuations), may be coupled leading to feedback between different channels.

In addition to the $q=0$ PIs, a continuum of divergences occur for finite q within the TCSE method - we shall refer to these points in parameter space as ``generalized Pomeranchuk instabilities" (GPIs). This paper will also study the physics of systems in the vicinity of GPIs in the $\ell=0$ channel.

In contrast to the work discussed above~\cite{ChubukovCharge, ChubukovSpin} and previous work using TCSE method \cite{QBB87}, in which PIs are approached starting from the paramagnetic state, the starting point of this work is the ferromagnetic state.
For this we use the well-established ferromagnetic Fermi liquid theory of Abrikosov, Dzyaloshinskii, and Kondratenko \cite{AD,DK}, valid for weakly ferromagnetic systems. Starting in the magnetically ordered state, and for an underlying zero-range interaction, we find several interesting results: Both $\ell=0$ ferromagnetic and charge density PIs are approached simultaneously, thereby displaying ``multicritical" behavior in these branches of solutions. Additionally, upon approach to these PIs, the system crosses a d-wave nematic PI in the charge channel, i.e. $F_{2}^{s}\rightarrow -5$, suggesting that a nematic transition may precede the FM and density instabilities.  What is remarkable is the  emergence of a nematic transition even with a zero-range interaction, and the inclusion of only the $\ell=0$ quantum fluctuations through  the TCSE scheme. These fluctuations are sufficiently long-ranged to give information about 
exchange of fluctuations in  higher angular momentum channels. We also explore pairing instability near PIs. We find that in the FM state, both singlet and triplet pairings are possible for a repulsive driving interaction, though singlet pairing is favored. This raises the intriguing possibility of switching between singlet and triplet pairing via some symmetry-breaking mechanism.

\section{TCSE Method} 
\label{sec:method} 

Strong interactions are encountered in many interacting Fermi systems; perturbation theory using these underlying bare potentials may diverge at short-range, long-range, or both. To avoid these problems, renormalized interactions and full vertices need to be considered. One way this can be done is via the fermion parquet or fully crossing-symmetric approach \cite{JacksonLandeSmith, BickersScalapino, Parquet}, which is non-perturbative in that it sums 2-body planar diagrams to infinite order in the particle-partlcle (p-p), particle-hole (p-h), and exchange particle-hole (ex-p-h) channels; see Fig.~\ref{fig:parquet}. This full treatment is an arduous task. Microscopic treatment of the p-p channel (Brueckner theory) prevents only short-range divergences, and microscopic treatment of the p-h channels (RPA) prevents only long-range divergences; so the completely reducible two-body vertex must indeed include both these types of renormalizations.  This implies that a consistent Fermi liquid theory cannot be formulated in terms of short-range effective interactions alone; collective excitations generated by these interactions must be exchanged between quasiparticles. This underscores the physical basis for TCSE, which can be considered a ``minimal" or 
``tractable" parquet~\cite{QuaderLittlePaper}.

\begin{figure}[t!]
\centering
\includegraphics[scale=0.5]{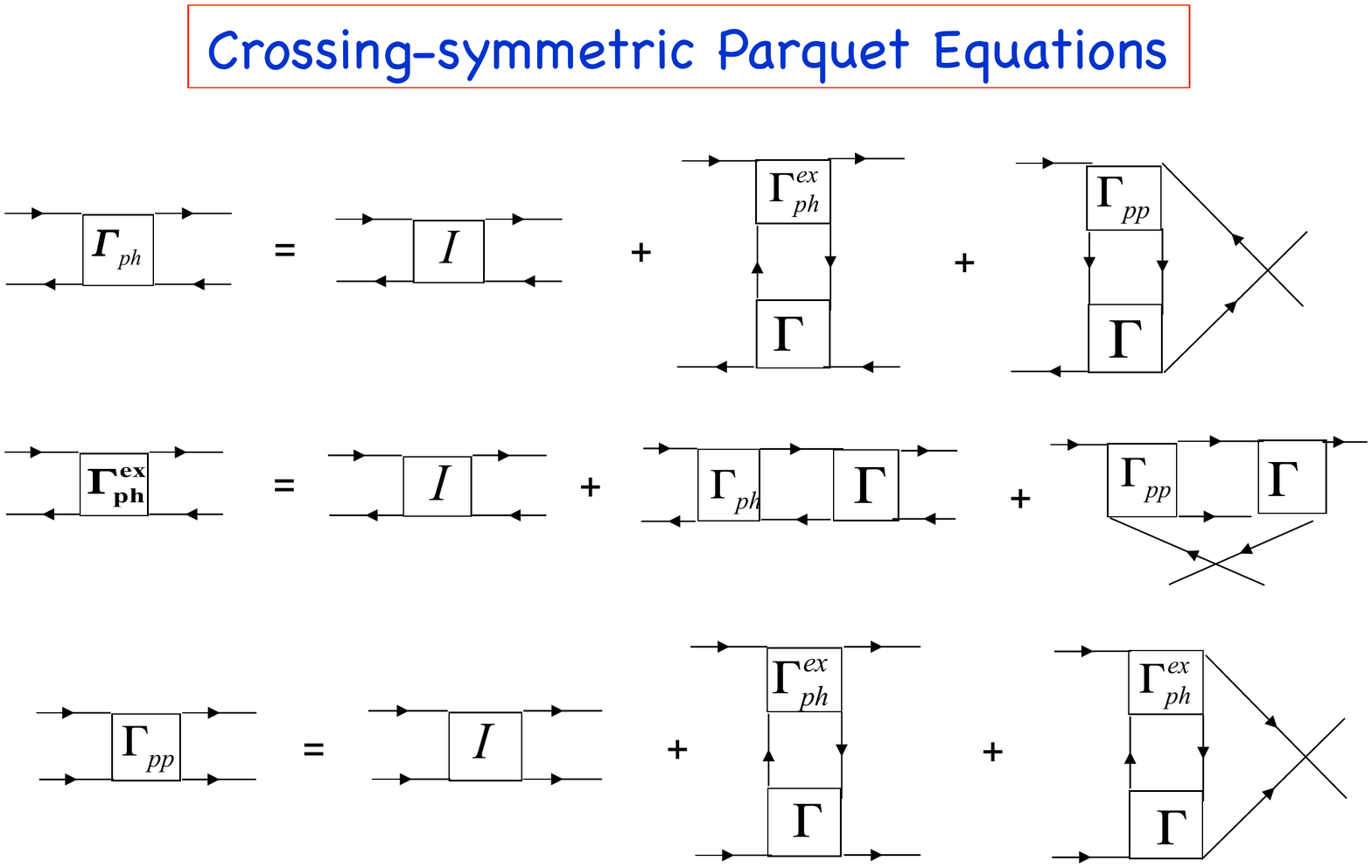}
\vspace{-4em}
\caption{Schematic form of crossing-symmetric parquet equations for the p-h, exchange p-h, and p-p vertex functions, $\Gamma_{ph}$.
$\Gamma_{ph}^{exch}$, and $\Gamma_{pp}$ respectively. $I$ represents completely irreducible vertex and $\Gamma$ the full 2-body vertex.}\label{fig:parquet}
\end{figure}

%\cite{JacksonLandeSmith, BabuBrown, BickersScalapino, AB87, ABBQ}. 

The fully crossing-symmetric  non-linear parquet equation for two-body vertex, $\Gamma$, can be formally written in terms of completely irreducible diagrams (I) and the two-body (p-p or p-h) Green's functions, $G_{i}$ ($i = s, t, u$ channel). 

\begin{equation} \Gamma = I+\displaystyle\sum_{i=s,t,u}\Gamma[I+G_{i}r]^{-1}G_{i}r\end{equation}
 
The TCSE method utilizes the idea that a large part of the renormalization of quasiparticle interaction comes through the p-h processes near the Fermi surface. This suggests the regrouping of diagrams into p-h reducible and irreducible terms. The \emph{tractable} crossing-symmetric equations are obtained via application of this regrouping along with partial re-summation of certain diagrams, quasiparticle renormalization, and careful preservation of crossing symmetry. Appropriate phase space is represented by Lindhard functions. In the TCSE scheme, these phase space functions are in terms of dressed Green's functions. TCSE are thus in terms of renormalized interactions in coupled particle-hole channels, in which these coupled channels feed back into each other.

The crossing-symmetric equations which result are a set of coupled non-linear integral equations for Landau interaction functions $F(q)$, and scattering amplitudes $A(q)$. The renormalized FL interaction functions is expressible in terms of a driving term and quantum fluctuation terms. The driving term, $D(q)$, contains diagrams that are particle-hole irreducible in both the direct (t) and exchange (u) particle-hole channels, such as all p-p terms (i.e. t-matrix and non-local interactions). It is model-dependent and reflects the symmetry of the underlying Hamiltonian; the choice of an antisymmetrized direct interaction is necessary to preserve the required crossing symmetry. Quantum fluctuation (QF) terms contain diagrams that are particle-hole reducible in the exchange p-h channel,
and accounts for medium effects and exchange of collective excitations, such as density, spin-density, and higher-order fluctuations. An important aspect of the TCSE method is its ability to treat an arbitrary underlying interaction ($D$) and these competing quantum fluctuations on the same footing. The set of equations obtained is shown schematically in Figure~\ref{fig:diagrams}. 

\begin{figure}[t]
\centering
\includegraphics[scale=0.5]{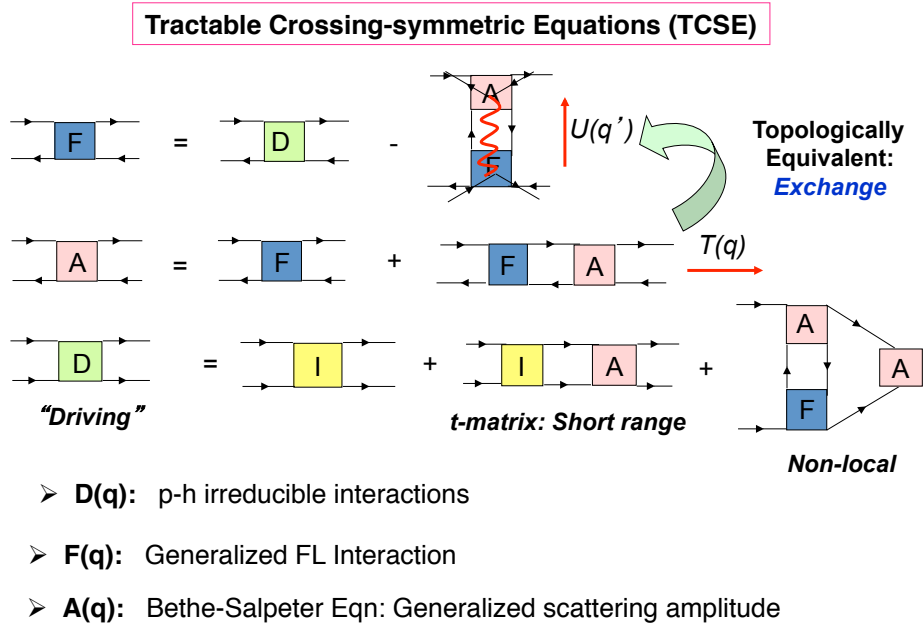}
\vspace{-3em}
\caption{Schematic form of tractable crossing-symmetric equations: $F$ (Landau interaction function), $A$ (scattering amplitude), $D$ (direct term), $I$ (completely irreducible diagrams). $T(q)$ and $U(q^{\prime}$) represent p-h and exchange p-h channels and are exchanges of each other.}\label{fig:diagrams}
\end{figure}

In a system with two species of spins, the FL interaction functions can be expressed as a combination of spin-symmetric (s)  and spin-antisymmetric (a) terms: 
 \begin{equation}F_{pp'}^{\sigma\sigma'}= F_{pp'}^{s}+F_{pp'}^{a}\vec{\sigma}\cdot\vec{\sigma'}
 \end{equation} 
 These interaction functions are expanded in Legendre polynomials for the 3D isotropic FL case, and can be calculated within the TCSE scheme. The FL functions are scaled to the density of states $N(0)$ at the Fermi surface, i.e. $F_\ell^{s,a}=N(0)f_\ell^{s,a}$ and $F_{pp'}^{s,a}=N(0)f_{pp'}^{s,a}$. $q'$ is the momentum transfer in the \emph{exchange} p-h channel;  similarly, q is the momentum transfer in the \emph{direct} particle-hole channel. Near the Fermi surface, $q'^2=2k_{F}(1-\cos\theta_L)$, where $\theta_L=\hat{p}\cdot\hat{p'}$; likewise for $q$.
 The FL interaction functions and direct (driving) interactions can be expanded in Legendre polynomials as
\begin{equation}\begin{split}
	F_{pp'}^{s,a}=\displaystyle\sum_{\ell}F_{\ell}^{s,a}P_{\ell}(\cos\theta_L) \\
	D_{pp'}^{s,a}=\displaystyle\sum_{\ell}F_{\ell}^{s,a}P_{\ell}(\cos\theta_L)
\end{split}\end{equation}	

Then the TCSE equation are then given by:
 
\begin{multline}
F_{pp'}^{s}=D_{pp'}^{s}+\frac{1}{2}\frac{F_{0}^{s}(q')\chi_0(q)F_{0}^{s}(q')}{1+F_{0}^{s}(q')\chi_0(q')}+\frac{3}{2}\frac{F_{0}^{a}(q')\chi_0(q)F_{0}^{a}(q')}{1+F_{0}^{a}(q')\chi_0(q')} \\ +\frac{1}{2}\left[1-\frac{q'^2}{4k_{F}^2}\right]\left[\frac{F_{1}^{s}\chi_{1}(q')F_{1}^{s}}{1+F_{1}^{s}\chi_{1}(q')}+3\frac{F_{1}^{a}\chi_{1}(q')F_{1}^{a}}{1+F_{1}^{a}\chi_{1}(q')}\right]+\ell=2,3\ldots 
\end{multline}
\begin{multline}
F_{pp'}^{a}=D_{pp'}^{a}+\frac{1}{2}\frac{F_{0}^{s}(q')\chi_0(q')F_{0}^{s}(q')}{1+F_{0}^{s}(q')\chi_0(q')}-\frac{1}{2}\frac{F_{0}^{a}(q')\chi_0(q')F_{0}^{a}(q')}{1+F_{0}^{a}(q')\chi_0(q)} \\+\frac{1}{2}\left[1-\frac{q'^2}{4k_{F}^2}\right]\left[\frac{F_{1}^{s}\chi_{1}(q')F_{1}^{s}}{1+F_{1}^{s}\chi_{1}(q')}-\frac{F_{1}^{a}\chi_{1}(q')F_{1}^{a}}{1+F_{1}^{a}\chi_{1}(q')}\right]+\ell=2,3\ldots
\end{multline}

where $\chi_{0}(q')$ and $\chi_{1}(q')$ are Lindhard functions -- density-density and current-current correlation functions  respectively; In 3D these are given by~\cite{FetterWalecka}: 

\begin{align}
&\chi_{0}(q')=\frac{1}{2}\left[1+\left(\frac{q'}{4}-\frac{1}{q'}\right)ln\Bigl|\frac{1-0.5q'}{1+0.5q}\Bigr|\right] \\
&\chi_{1}(q')=\frac{1}{2}\left[\frac{3}{8}-\frac{1}{2q'^2}-\left(\frac{1}{2q'^3}+\frac{1}{4q'}-\frac{3q'}{32}\right)ln\Bigl|\frac{1-0.5q'}{1+0.5q}\Bigr|\right]
\end{align}
 
For a given driving interaction, $D$, the FL interaction functions can be calculated in any angular momentum channel, along with the corresponding scattering amplitudes and effective mass (related to self-energy). From these basic quantities, various transport, thermodynamic, and pairing properties can then be calculated.

\section{Model}
\label{sec:model}

For our model calculations here, we choose a zero-range driving interaction, as in the single-band Hubbard model. 
This model is one of the most-studied in correlated electron systems in condensed matter physics. With this interaction as the direct term, the antisymmetrized $\ell=0$ direct term takes the form
\begin{equation}
D_{0}^{s,a}=\pm\frac{U}{2}
\end{equation}
where $U$ is scaled to density of states at the Fermi surface.

For the case in which only the $\ell=0$ density and spin fluctuations are included, and  the FL interaction functions themselves have no explicit momentum-dependence, the model TCSE coupled non-linear integral equations are given by:
	
\begin{equation}\begin{split}
F_{0}^{s}=\frac{U}{2}&+\frac{1}{2}\int_0^{2k_F} \! \frac{F_0^{s}\chi_0(q')F_0^{s}}{1+F_0^{s}\chi_0(q')} \, \mathrm{d} q'+\frac{3}{2}\int_0^{2k_F }\! \frac{F_0^{a}\chi_0(q')F_0^{a}}{1+F_0^{a}\chi_0(q')} \, \mathrm{d} q' \\
F_{0}^{a}=\frac{-U}{2}&+\frac{1}{2}\int_0^{2k_F }\! \frac{F_0^{s}\chi_0(q')F_0^{s}}{1+F_0^{s}\chi_0(q')} \, \mathrm{d} q'-\frac{1}{2}\int_0^{2k_F} \! \frac{F_0^{a}\chi_0(q')F_0^{a}}{1+F_0^{a}\chi_0(q')} \, \mathrm{d} q'
\end{split}\end{equation}
\\

\section{Solutions}
\label{sec:solution techniques}

\subsection{Parameter space for generalized Pomeranchuk Instabilities}
\label{sec:parameter space}

A goal of this work is to perform calculations starting not only from the disordered paramagnetic phase, but also starting from 
the ordered ferromagnetic phase. 
On the ordered side, our starting ground state is the weak ferromagnetic FL, based on the well-established theory of Abrikosov, Dzyaloshinskii, and Kondratenko \cite{AD,DK}. Since the TCSE equations give $F_{\ell}^{s,a}$ as solutions, we can 
depict a parameter space defined by  $F_{0}^a,F_{0}^s$. The space can be thought of as comprising of four regions: ferromagnetic, paramagnetic, phase separated, and mixed phase (meaning both charge and ferromagnetic PI thresholds have been crossed, so both types of instabilities are present). See Table~\ref{RegionTable} and  Fig.~\ref{Regions} which shows the $F_{0}^a,F_{0}^s$ parameter space. The thatched sections of the figure are the regions of finite-q divergences, bounded on one edge by the $q=0$ PI ($F_{0}^{s,a}\rightarrow -1$)and on the other edge by a $q=2k_F$ instability ($F_{0}^{s,a}\rightarrow -2$), with all other generalized Pomeranchuk instabilities (GPI) between these boundaries.
As will be seen later, existence and nature of solutions in different regions of phase space depend 
on whether the underlying interaction (given by the driving term $D$) is repulsive or attractive.\\

\begin{table}
\centering
\begin{tabular} { l    c    c    c    c   }
\hline
Solution Sector       &    \\ \hline
Ferromagnetic  & $F_{0}^{a}<-1, F_{0}^{s}>-1$  \\
Paramagnetic  & $F_{0}^{a}>-1, F_{0}^{s}>-1$   \\
Mixed (FM/CDW)  &  $F_{0}^{a}<-1, F_{0}^{s}<-1$ \\
Phase separation  &  $F_{0}^{a}>-1, F_{0}^{s}<-1$ \\
\hline
\end{tabular}
\caption{Solution sectors as defined in parameter space.}
\label{RegionTable}
\end{table}

\begin{figure}[h!]
\center{\includegraphics[scale=0.4]{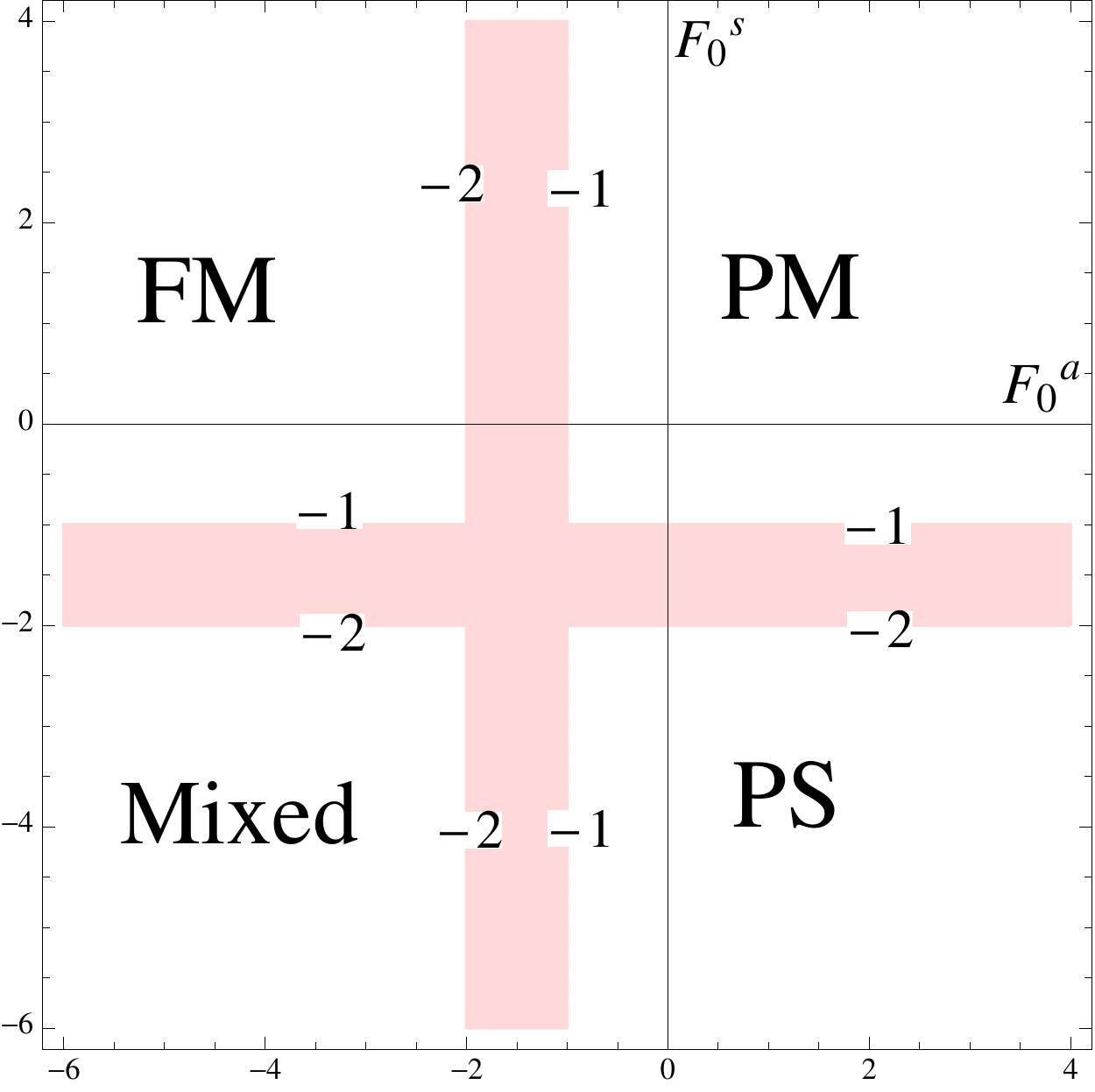}}
\caption{Regions of $F_{0}^{a}, F_{0}^{s}$ parameter space and generalized Pomeranchuk instabilities.}\label{Regions}
\end{figure}

\subsection{Graphical and numerical methods}
\label{sec:techniques}

To find solutions to the TCSE,  we employ both graphical and numerical techniques. In conjunction with numerical evaluation of integrals, graphical techniques are utilized to find self-consistent solutions to the set of coupled non-linear TCSE above.
To solve these coupled equations, it is convenient to cast the expressions in the spin-symmetric (S) and antisymmetric (A) channels as
follows: 

\begin{equation}\begin{split}
S=F_{0}^{s}-\frac{U}{2}-\frac{1}{2}\int_0^{2k_F} \! \frac{F_0^{s}\chi_0(q')F_0^{s}}{1+F_0^{s}\chi_0(q')} \, \mathrm{d} q'-\frac{3}{2}\int_0^{2k_F }\! \frac{F_0^{a}\chi_0(q')F_0^{a}}{1+F_0^{a}\chi_0(q')} \, \mathrm{d} q' \\
A=F_{0}^{a}+\frac{U}{2}-\frac{1}{2}\int_0^{2k_F }\! \frac{F_0^{s}\chi_0(q')F_0^{s}}{1+F_0^{s}\chi_0(q')} \, \mathrm{d} q'+\frac{1}{2}\int_0^{2k_F} \! \frac{F_0^{a}\chi_0(q')F_0^{a}}{1+F_0^{a}\chi_0(q')} \, \mathrm{d} q'
\end{split}\end{equation}

When both equations are satisfied, that is,
\begin{equation}
S=A=0
\end{equation}
the corresponding point ($F_{0}^a$, $F_{0}^s$) in parameter space is a solution to the TCSE.
Thus, solutions can be found graphically as the three-way intersection of the symmetric channel equation (S), the antisymmetric channel equation (A), and the zero plane in this three-dimensional space. An example of a graphical solution plotted in these three dimensions can be seen in Fig.~\ref{Pole}. This can also be seen (not shown) in a two-dimensional plot in which the space is the zero plane slice of three-dimensional space discussed above. Then the surfaces S and A appear as lines, and their intersection represents a solution. 
 
Once the $\ell=0$ interaction functions have been calculated using methods discussed,  FL interaction functions in higher angular momentum channels are obtained using the orthogonality of the Legendre polynomials to project out the FL interaction functions in any desired channel. 
The scattering amplitudes are calculated from the $F_\ell$'s as \begin{equation} A_{\ell}^{s,a}=\frac{F_{\ell}^{s,a}}{1+F_{\ell}^{s,a}/(2\ell+1)}\end{equation} 
One can use Landau forward scattering sum rule to check the convergence of these scattering amplitudes: 
 $\displaystyle\sum_{\ell}(A_\ell^s+A_\ell^a)=0$. 

Superconducting pairing amplitudes (singlet and triplet amplitudes, $g_{s,t}$) are calculated using the Patton-Zaringhalam scheme~\cite{PZ},  giving  
\begin{equation}\label{eq:pz}\begin{split}
\textrm{Singlet:   } g_s=\sum_{\ell} (-)^{\ell}(A_{\ell}^{s}-3A_{\ell}^{a})/4
 \\
\textrm{Triplet:   } g_t=\sum_{\ell} (-)^{\ell}(A_{\ell}^{s}+A_{\ell}^{a})/12
\end{split}\end{equation}

\subsection{Finite-q divergences: Generalized Pomeranchuk instabilities (GPIs)}
\label{sec:Extension to finite q}

The momentum-dependence of  the phase space introduces finite-momentum divergences to the problem, in addition to the $q$=0 PIs. These finite-q divergences were described in the sub-section on "generalized Pomeranchuk instabilities" (GPI) and are shown in the parameter space of Fig.~\ref{Regions}. The quantum fluctuation terms diverge when $1+F_{0}^{s,a}\chi_0(q')=0$. Since $0.5 \leq \chi_0(q') \leq 1$ for $0 \leq q \leq 2k_F$.  When $-2 \leq F_{0}^{s,a} \leq -1$, there are two divergences for \emph{every} value of $q^\prime$: one in the $F_{0}^a$ integral and one in the $F_{0}^s$ integral. So, there exist two uncountably infinite sets of divergences. Numerically, for every value of $q^\prime$ sampled in the integration, one divergence is present in each of the two ($s,a$) channels. These sets of divergences are bounded in parameter space for $\ell=0$ by $F_{0}=-2$ on the negative edge and $F_{0}=-1$ on the positive edge (see shaded regions in Fig.~\ref{Regions}). We treat these sets of divergences using a numerical contour integration, as shown below.

\begin{equation}
\int_0^{2k_F} \! \frac{\chi_0(q')}{1+F_0^{s,a}\chi_0(q')} \, \mathrm{d} q' \rightarrow \int_\pi^0 \! i\epsilon e^{ix}\frac{\chi_0(q_{pole}+\epsilon e^{ix})}{1+F_0^{s,a}\chi_0(q_{pole}+\epsilon e^{ix})}\, \mathrm{d} x
\end{equation}
\\
\begin{figure}[h!]
\centering
\includegraphics[scale=0.4]{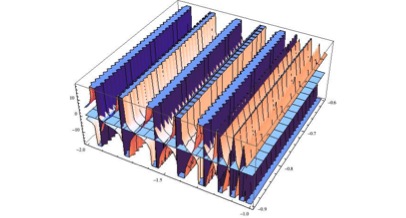}
\includegraphics[scale=0.4]{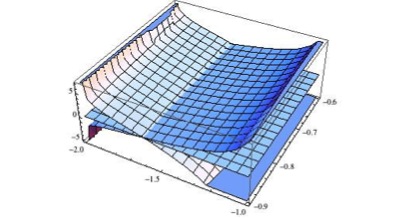}
\caption{$U$=12 equation surfaces (left) prior to, and (b) after contour integration around divergences.}\label{Pole}
\end{figure}

The surfaces defined by the equations are shown in Fig.~\ref{Pole} before and after the treatment of poles via numerical contour integration. The surfaces shown are the TCSE as functions of $F_{0}^{s,a}$ when $U=12$. In Fig.~\ref{Pole} (before contour integration), the spikes along the $F_{0}^a$ axis are poles at Gauss points sampled in the integration of $F_{0}^a$ in the quantum fluctuation terms of the TCSE's. The more Gauss points used, the more spikes appear in the surfaces, since each Gauss point corresponds to a different value of $q$; recall that for each value of $q$, there is a divergence in each channel. If $F_{0}^s$ had also been plotted in the range of finite $q$ divergences ($F_{0}^s$ between -1 and -2), similar spikes would be present along its axis as well. It can be seen from Fig.~\ref{Pole} that the use of a numerical contour integration treats the finite $q$ divergences and makes them integrable in these regions. 

\section{Results}
\label{sec:results}

Solutions to Eq. (9) give $F_0^{s,a}$; then higher-order FL parameters are projected out. These are then used to obtain scattering amplitudes, effective mass, and pairing amplitudes. Since the TCSE  are non-linear, we find three primary branches of solutions, corresponding to ground states of different physical systems. The solution sectors in which these three solutions are found vary with the sign of the interaction. 

The three types of solutions found for the case of repulsive interactions include one paramagnetic and two ferromagnetic branches: one near the FM PI (weak FM) and one far beyond the FM PI (strong FM). At least one other branch of solutions is found, but guided by the behavior of solutions in the local FL limit and previous work \cite{EB_PM}, we deem these solutions unphysical and focus on the three branches mentioned; the extra solutions may be considered purely mathematical in natural, arising as a result of the nonlinearity of the TCSE equations. Considering the large-$U$ limit of the solutions gives insight into the general behavior and properties of these solutions. See Table~\ref{LargeURepTable} for these limiting results in the general (non-local) case of a repulsive contact interaction.

\begin{table}[b]
\centering
\begin{tabular} { l    c    c    c    c   }
\hline
Solution Type       &    $F_{0}^{a}$ & $F_{0}^{s}$ &  $(m*/m)$ & Pairing \\ \hline
PM  & $-0.63$ & $U$ & Large & Triplet \\
FM (strong)  & $-U/3$ & $-0.63$ & Modest & Singlet/Triplet  \\
FM (weak)  &  $-2^{(-)}$ & $-1^{(+)}$ & Large & Singlet/Triplet  \\
\hline
\end{tabular}
\caption{Large-$U$ limiting results for repulsive contact interaction.}\label{LargeURepTable}
\end{table}

In the paramagnetic branch, in agreement with previous work \cite{QBB87, EB_PM},  we find that $F_{0}^{a}$ approaches $-0.63$ from the positive side, meaning it moves in the direction of the FM instability with increasing $U$, and $F_{0}^{s}$ approaches $U$. The effective mass (related to $F_{1}^{s}$) is large and of the order of $U$. Using Patton-Zaringhalam scheme~\cite{PZ} (Eq.~\eqref{eq:pz}) to calculate singlet and triplet effective pairing amplitudes, we find that only triplet pairing is found to have an attractive amplitude in this branch. 

The next solution branch is the strongly ferromagnetic solution. Here, $F_{0}^{a}$ approaches $-U/3$ and $F_{0}^{s}$ approaches $-0.63$ from the positive side (moving in the direction of the density instability with increasing $U$). Note that as the underlying repulsive interaction increases, this solution moves deeper into the ferromagnetic regime. While weakly FM systems may be described by Abrikosov-Dzyaloshinskii-Kondratenko FM Fermi liquid theory \cite{AD, DK}, we note that no rigorous Fermi liquid theory exists at present for describing strongly FM systems. In this branch, the effective mass is modest but greater than one. In contrast to the PM solution, both singlet and triplet pairing amplitudes are attractive (and thus both types of pairing are possible), but singlet is favored due to its larger negative magnitude. 

\begin{figure}
\centering
\includegraphics[scale=0.4]{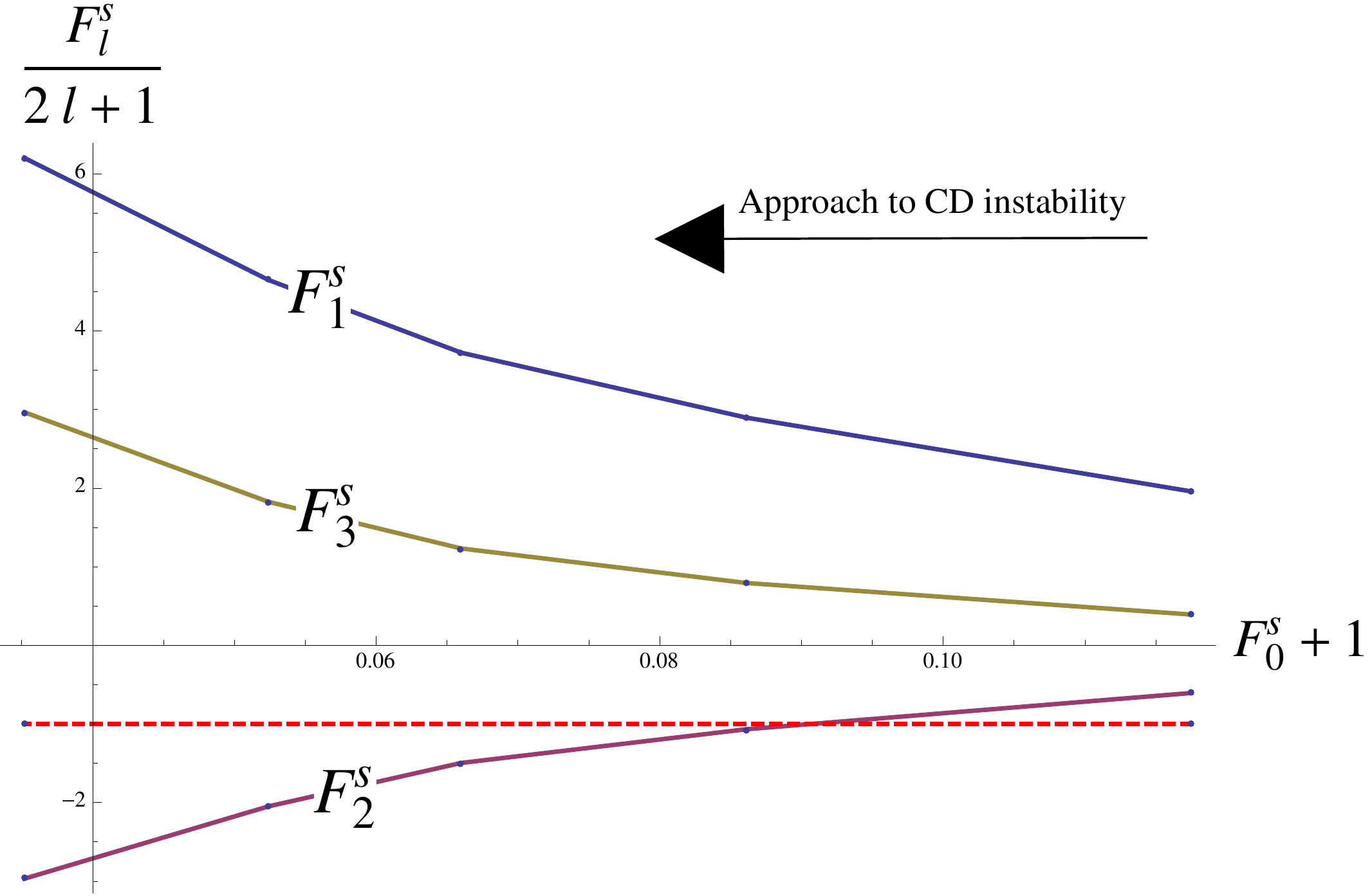}
\caption{Scaled spin-symmetric Landau parameters ($F_{\ell}^{s}$) upon approach to charge density instability. Dashed line indicates position of $q'=0$ PI for any channel. Note that $F_{2}^{s}$ crosses its instability ($\frac{F_{2}^{s}}{5}=-1$) before $F_{0}^{s}$ reaches -1 (the CD instability).}\label{ChargeToChargePI}
\end{figure}

 \begin{figure}
 \centering
\includegraphics[scale=0.4]{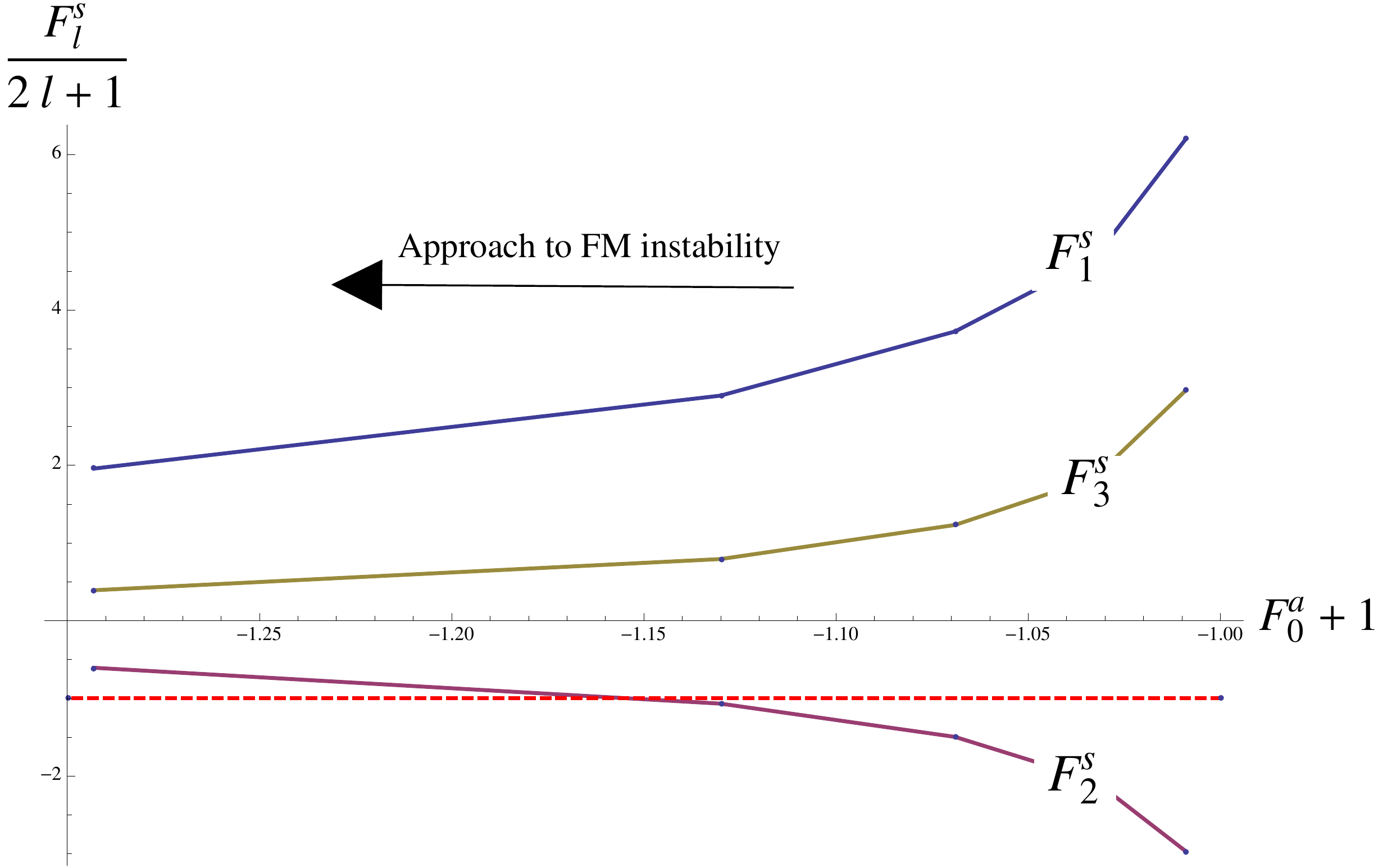}
\caption{Scaled spin-symmetric Landau parameters ($F_{\ell}^{s}$) upon approach to FM instability. Dashed line indicates position of $q'=0$ PI for any channel. Note that $F_{2}^{s}$ crosses its instability ($\frac{F_{2}^{s}}{5}=-1$) before $F_{0}^{a}$ reaches -2 (the GPI associated with the  FM instability).}\label{ChargeToFMPI}
\end{figure}

Finally, in the weakly ferromagnetic branch that is near the FM PI, $F_{0}^{a}$ approaches $-2$ from the negative side and $F_{0}^{s}$ approaches $-1$ from the positive side. Due to the region of GPI's in the antisymmetric channel, $F_{0}^{a}$ cannot move all the way to the FM instability, but instead gets ``stuck" at the $q'=2$ GPI of -2 (possible SDW instability). 
In calculations not reported here, we have found that with explicit $q$-dependence in the FL interactions, $F_{0}^{a}$ 
does approach the FM ($q'=0$) PI of -1 with increasing $U$. Of particular significance in this weak FM branch is the emergence of ``multicritical" behavior in the systems,  meaning that both spin and density channels move towards instabilities as $U$ increases.
The weak FM branch of solutions is the only branch found to exhibit this quantum multicritical behavior. The effective mass in this weak FM branch is found to be large and of the order of $U$. As in the other FM solution branch, both singlet and triplet pairing are attractive, but singlet pairing is preferred. 

\subsection{Nematic instability}

For repulsive interactions, the weak FM branch of solutions exhibits another fascinating behavior in addition to its multicriticality. Upon approach to either the FM or CD instability, $\chi_2(q^\prime)$ diverges in the spin-symmetric channel, leading to a charge nematic instability. In FL language, $F_{2}^{s}\rightarrow-5$.  Thus a charge nematic transition both precedes and is driven by the approach to the s-wave instabilities. As note above, our calculations show that the system approaches the $q=0$ FM instability with $q^\prime$-dependence in $F_0^a$, but even without this $q$-dependence, $F_{2}^{s}$ approaches its PI. Fig.~\ref{ChargeToChargePI} and ~\ref{ChargeToFMPI} show the higher angular momentum harmonics of the FL interaction function in the charge channel upon approach to the FM and density instabilities, respectively, with no $q'$-dependence added to $F_{0}^{a}$. Fig.~\ref{SpinChargePI} show the higher angular momentum harmonics of the FL interaction function in the spin channel upon approach to the FM and density instabilities, respectively, also with no $q'$-dependence added to $F_{0}^{a}$.

\begin{figure}
 \centering
\includegraphics[scale=0.27]{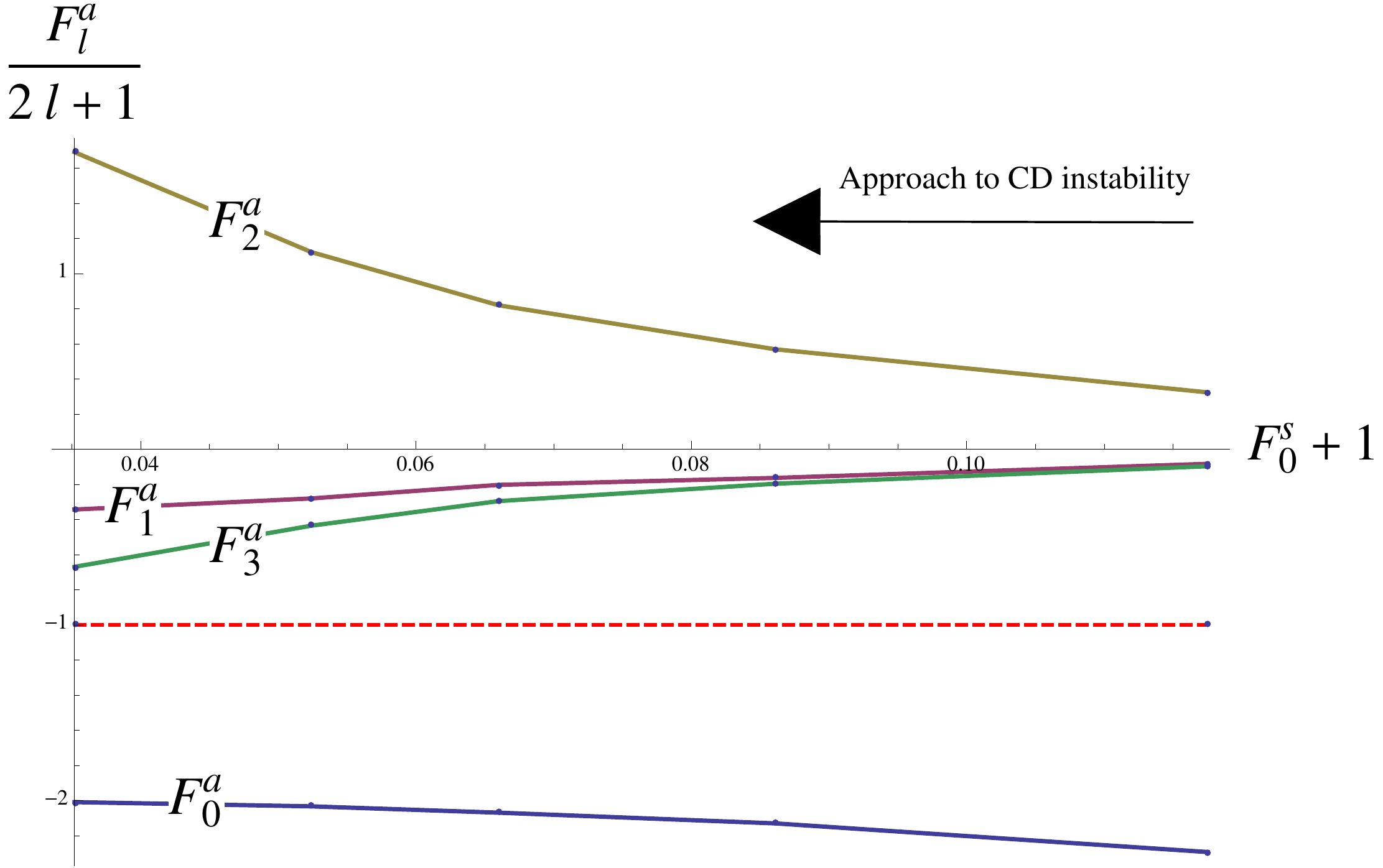}
\includegraphics[scale=0.27]{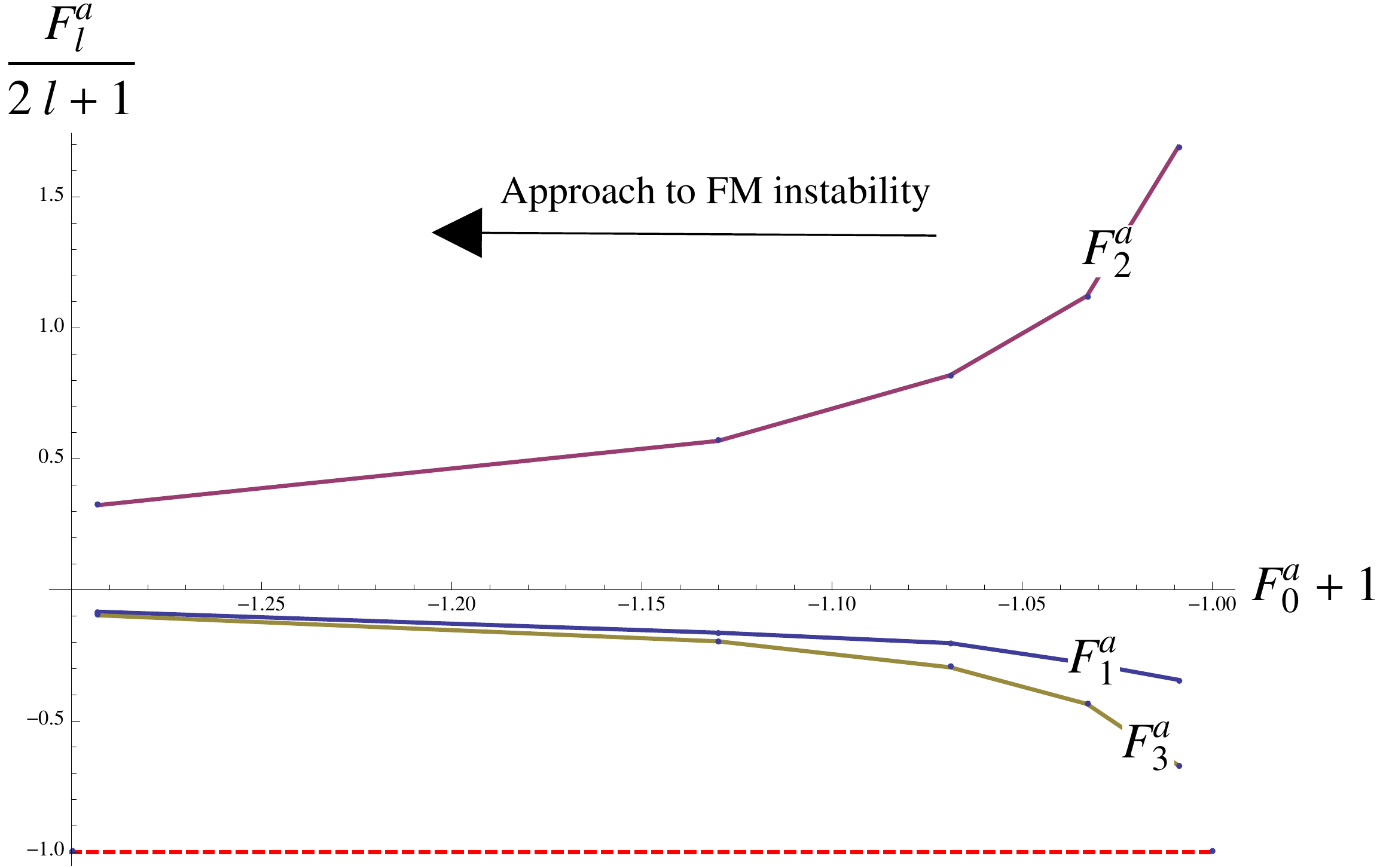}
\caption{(left) Scaled spin-antisymmetric Landau parameters ($F_{\ell}^{a}$) upon approach to charge density instability. 
(Right) Scaled spin-antisymmetric Landau parameters ($F_{\ell}^{a}$) upon approach to FM
instability. Dashed lines indicates position of $q'=0$ PI for any channel. Note that no $F_{\ell}^{a}$s reach an instability before 
$F_{0}^{s}$ reaches -1, or before $F_{0}^{a}$ reaches -2.}~\label{SpinChargePI}
\end{figure}
	
\section{Summary and discussion} 
\label{sec:summ} 

In the TCSE method, an arbitrary underlying interaction is treated on the same footing as quantum fluctuations. In our model, only $\ell=0$ fluctuations are taken into account. $F_{0}^{s,a}$ can then be thought of as a sum of the direct interaction and quantum fluctuations. 
Upon approach to PIs's (with increasing $U$), quantum fluctuations are enhanced in both strong and weak (multicritical) FM branches of solutions,  as evident from discussions and figures above. In the spin channel, fluctuations work together to cancel most of the contribution from the driving interaction, which leads to a small $F_{0}^{a}$, where by ``small", we mean ``close to $F_{0}^{a}$ approaching the Pomeranchuk instability. 
In FM FL theory, ferromagnetism is related to $F_{0}^{a}$ as $m\sim[1+F_{0}^{a}]^{1/2}$. In the multicritical branch of solutions, this quantity becomes weaker for increasing direct interaction, as opposed to becoming stronger, as in the strongly FM branch. 
This counterintuitive behavior of $F_{0}^{a}$ is due to the large fluctuations which are opposite in sign to the direct interaction, leading to a small $F_{0}^{a}$, which gets even smaller for increasing $U$ (i.e. approach to PI) due to enhancement of fluctuations in this area.
In the density channel, the density fluctuations, together with the driving interaction, compete against spin fluctuations to give a small $F_{0}^{s}$, where by ``small", we mean close to the $F_{0}^{s}$ PI. Thus, this enhancement, interplay, and feedback of quantum fluctuations results in multicritical behavior in which both channels simultaneously approach a GPI. We have found that with explicit $q$-dependence in $F_{0}^{a}$, which allows $F_{0}^{a}$ to approach its $q=0$ PI ($F_{0}^{a}=-1$), the fluctuations in the density channel do not change qualitatively. However, the spin fluctuations in the spin channel are stronger and compete more with density fluctuations. 

Fluctuations in spin and density channels also affect the higher $\ell$ FL parameters, as evidenced by our discussions on
approach to nematic PI, effective mass and pairing amplitudes. 
Pairing amplitudes for the ferromagnetic solutions are found to be attractive for both singlet and triplet, but singlet is found to be more attractive. This is due to the interplay and competition between quantum fluctuations and direct interaction. This result raises the intriguing possibility of switching between singlet and triplet via some symmetry-breaking effect.

\section*{Acknowledgments} 
We thank Michael Widom, Andrey Chubukov, and Dimitri Maslov for useful discussions. We also acknowledge
the support of Institute for Complex Adaptive Matter (ICAM), and the hospitality
of Aspen Center for Physics, where part of the work was done.

\end{document}